\documentclass{article}

\usepackage{arxiv}

\usepackage[utf8]{inputenc} 
\usepackage[T1]{fontenc}    
\usepackage{hyperref}       
\usepackage{url}            
\usepackage{booktabs}       
\usepackage{amsfonts}       
\usepackage{nicefrac}       
\usepackage{microtype}      
\usepackage{lipsum}
\usepackage{graphicx}
\usepackage{amsmath}
\usepackage{array}
\usepackage{caption}
\graphicspath{ {./images/} }

\title{Engineering-Oriented Symbolic Regression: LLMs as Physics Agents for Discovery of Simulation-Ready Constitutive Laws}

\author{
  Yue Wu$^{a,*}$ \\ 
  $^{a}$Materials Genome Institute\\
  Shanghai University\\
  Shanghai, China 200444 \\
  \texttt{repeatsongyue@gmail.com} \\
  \And
  Tianhao Su$^{a,*}$ \\
  $^{a}$Materials Genome Institute\\
  Shanghai University\\
  Shanghai, China 200444 \\
  \texttt{thsu0407@gmail.com} \\
  \And
  Mingchuan Zhao$^{c}$ \\
  $^{c}$School of Mathematical Sciences\\
  University College Cork\\
  Cork, Republic of Ireland \\
  \texttt{mzhao@ucc.ie} \\
  \And
  Shunbo Hu$^{a,b,\dagger}$ \\
  $^{a}$Materials Genome Institute\\
  $^{b}$Institute for the Conservation of Cultural Heritage, \\
  School of Cultural Heritage and Information Management,\\
  Shanghai University\\
  Shanghai, China 200444 \\
  \texttt{shunbohu@shu.edu.cn} \\
  \AND
  Deng Pan$^{a,\dagger}$\\
  $^{a}$Materials Genome Institute\\
  Shanghai University\\
  Shanghai, China 200444 \\
  \texttt{DPan\_MGI@shu.edu.cn} \\
}
\date{}
\begin{document}
\maketitle 
\begingroup 
    \renewcommand{\thefootnote}{\fnsymbol{footnote}}
    \footnotetext[1]{These authors contributed equally to this work.}
    \footnotetext[2]{Corresponding author.}
\endgroup

\date{\vspace{-4ex}}

\maketitle
\thispagestyle{empty}
\thispagestyle{plain} 
\begin{abstract}
The discovery of constitutive laws for complex materials has historically faced a dichotomy between high-fidelity data-driven approaches, which demand prohibitive full-field experimental data, and traditional engineering fitting, which often yields numerically unstable models outside calibration regimes. In this work, we propose an Engineering-Oriented Symbolic Regression (EO-SR) framework that bridges this gap by leveraging Large Language Models (LLMs) as "Physics-Informed Agents." Unlike unconstrained symbolic regression, our framework utilizes an LLM Agent to zero-shot synthesize executable physical constraints—specifically thermodynamic consistency and frame indifference—transforming the search process from mathematical curve-fitting into a physics-governed discovery engine. We validate this approach on the hyperelastic modeling of rubber-like materials using standard Treloar datasets. The framework autonomously identifies a novel hybrid constitutive law that combines a Mooney-Rivlin linear base with a rational locking term. This discovered model not only achieves high predictive accuracy across multi-axial deformation modes (including zero-shot prediction of pure shear) but also guarantees unconditional convexity. Finite element validation demonstrates that while industry-standard models (e.g., Ogden N=3) fail due to numerical singularities under severe transverse compression, the EO-SR-discovered model maintains robust convergence. This study establishes a generalized, low-barrier pathway for discovering simulation-ready constitutive closures that satisfy both data accuracy and rigorous physical laws.
\end{abstract}


\begin{figure}[h!] 
    \centering
    \includegraphics[width=0.9\linewidth]{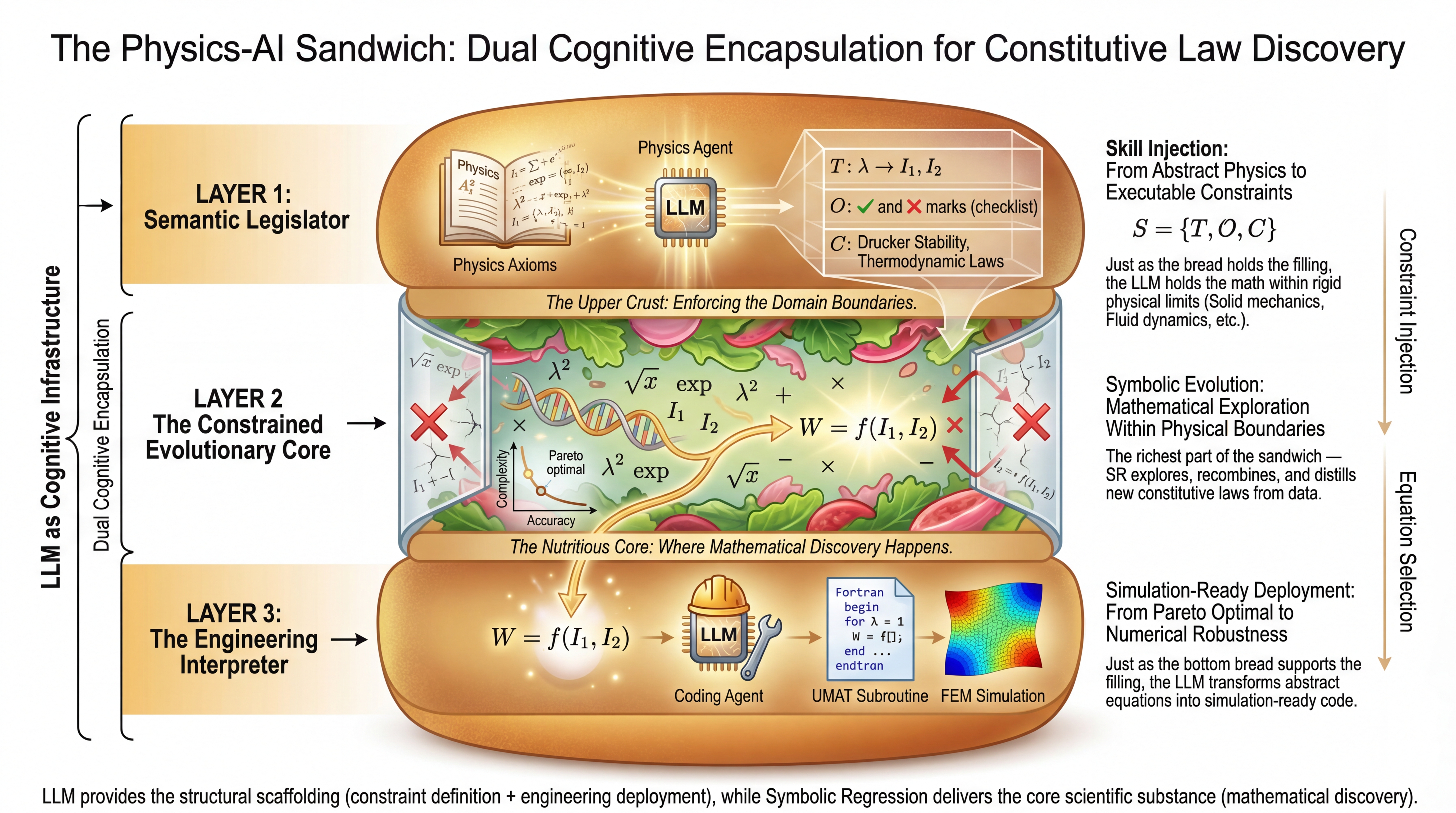} 
    \label{fig:toc}
\end{figure}

\section{Introduction}
The mathematical description of complex physical systems relies on two fundamental pillars: universal conservation laws (e.g., conservation of mass, momentum, and energy)\cite{truesdell1965non, gurtin1982introduction} and material-specific constitutive relations\cite{holzapfel2002nonlinear}. While the former are established from first principles, the latter—serving as the mathematical closures that define how matter interacts with external fields—remain the primary source of epistemic uncertainty in computational physics\cite{nath2024application}. From the stress-strain response in solid mechanics to turbulent closure models in fluid dynamics and flux-gradient relations in thermodynamics, the discovery of accurate, interpretable, and mathematically robust constitutive laws stands as a grand challenge across scientific disciplines\cite{wang2023scientific}.

In the era of ``AI for Science,'' the paradigm of constitutive modeling is undergoing a profound transformation\cite{karniadakis2021physics, linka2023new}. Data-driven approaches, particularly deep learning\cite{raissi2019physics} and symbolic regression\cite{schmidt2009distilling}, offer the potential to automate the discovery of these laws from experimental observations. However, a critical ``validity gap'' persists across these domains. A constitutive model is not merely a regression curve that minimizes fitting error; it must strictly adhere to an intrinsic set of physical constraints—such as thermodynamic consistency\cite{10.1115/1.4011929, vlassis2021sobolev} (e.g., entropy production inequality), frame indifference (objectivity), and mathematical stability\cite{marsden1994mathematical} (e.g., ellipticity or convexity). Neglecting these constraints allows purely data-driven algorithms to generate ``physically invalid'' expressions—models that may fit training data with high precision but violate fundamental laws of physics\cite{flaschel2021unsupervised}, leading to non-physical predictions or catastrophic convergence failures in downstream numerical simulations.

Historically, the discovery of constitutive laws has bifurcated into two distinct paradigms, each with inherent limitations regarding industrial applicability.

The first paradigm, often termed the ``\textbf{High-Fidelity Data-Driven}'' approach, seeks to extract constitutive manifolds from rich, heterogeneous data sources. Techniques utilizing Digital Image Correlation (DIC) combined with the Virtual Fields Method (VFM)\cite{pierron2012virtual, avril2008overview} or Physics-Informed Neural Networks (PINNs)\cite{karniadakis2021physics, raissi2019physics} have achieved remarkable success in capturing complex material behaviors without assuming specific analytical forms. However, this approach inherently raises the barrier to entry. It demands expensive experimental setups and specialized data processing pipelines that are often inaccessible in standard engineering environments. Furthermore, the resulting models, often parameterized as ``black-box'' neural networks\cite{kirchdoerfer2016data}, can obscure the phenomenological nature of continuum mechanics, sacrificing interpretability for localized precision.

The second paradigm, the ``\textbf{Traditional Engineering Fitting}'' approach, relies on calibrating pre-defined analytical templates (e.g., Ogden\cite{ogden1972large}, Yeoh\cite{yeoh1993some}, Mooney-Rivlin\cite{mooney1940theory, rivlin1948large}) against standard uniaxial or biaxial tensile data. While accessible and widely integrated into commercial Finite Element (FE) software, this approach suffers from a critical blind spot: the decoupling of fitting accuracy from numerical robustness. As demonstrated in this study, models optimized solely for fitting error (MSE) frequently violate fundamental thermodynamic constraints, such as the Drucker stability criterion\cite{10.1115/1.4011929} outside the calibration range. This leads to the ``fitting-simulation gap''\cite{holzapfel2000continuum}, where a model that fits experimental curves perfectly causes immediate convergence failure in complex boundary value problems due to mathematical artifacts like non-convexity or spurious softening.

Ideally, an engineering-oriented solution should bridge this divide: it should operate on low-cost standard data while enforcing rigorous physical validity. Symbolic Regression (SR) offers a promising pathway\cite{schmidt2009distilling, cranmer2020discovering} by searching for free-form analytical equations. However, standard SR functions as an unconstrained mathematical search engine, prone to generating ``physically invalid'' expressions\cite{flaschel2021unsupervised}, equations that fit data but violate basic laws of physics.

\textbf{This is where the rapid ascent of Large Language Models (LLMs) offers a transformative solution.\cite{openai2023gpt, bubeck2023sparks, wolf2020transformers}} Beyond their traditional capabilities in natural language processing, modern foundation models have exhibited emergent abilities in scientific reasoning\cite{wei2022chain, boiko2023autonomous} and automated code synthesis\cite{chen2021evaluating, romera2024mathematical, surameery2023use}. They can now bridge the semantic gap between abstract physical principles (e.g., ``strain energy must be convex'') and executable mathematical logic. In this work, we propose an Engineering-Oriented Symbolic Regression (EO-SR) framework that leverages LLMs not as generative black boxes, but as ``Physics-Informed Agents''\cite{xi2025rise, m2024augmenting, wang2023voyager, 10.1145/3586183.3606763} to explicitly constrain the discovery process.

The core contributions of this paper are as follows:

\begin{enumerate}
    \item \textbf{A Skill-Based Discovery Architecture:} We introduce a modular framework where an LLM Agent acts as a domain expert, zero-shot generating executable physical constraints (e.g., convexity, frame indifference) that guide the symbolic evolution. This transforms SR from a mathematical curve-fitter into a physics-governed discovery engine.
    
    \item \textbf{Bridging the Fitting-Simulation Gap:} We demonstrate that by injecting thermodynamic constraints (specifically Drucker stability) into the loss function, the framework can discover constitutive laws from standard Treloar data\cite{treloar1944stress} that are unconditionally stable in Finite Element Analysis (FEA), eliminating the numerical risks associated with traditional fitting.
    
    \item \textbf{Discovery of Optimal Forms:} The framework identifies a novel hybrid constitutive model that combines the linearity of Mooney-Rivlin with a rational locking term\cite{gent1996new}. This discovered model outperforms industry standards (Ogden, Yeoh) in terms of generalization to pure shear and numerical robustness, validating the efficacy of the proposed methodology.
\end{enumerate}

\section{Methodology}
\subsection{The EO-SR Framework Architecture}
\begin{figure*}[htbp]
    \centering
    \includegraphics[width=\textwidth]{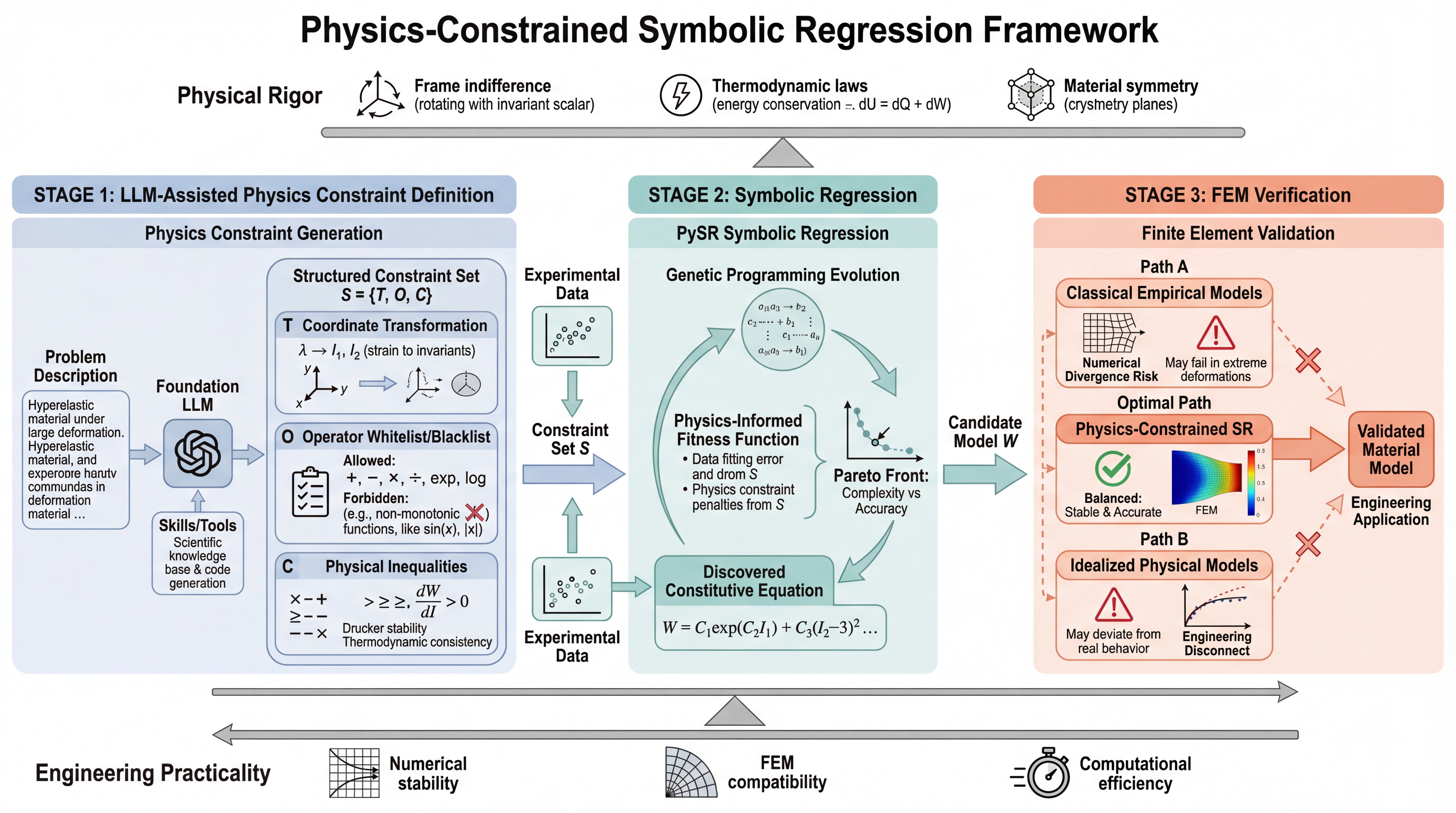} 
    \caption{\textbf{Overview of the Engineering-Oriented Symbolic Regression (EO-SR) framework.} The workflow is designed to bridge physical rigor with engineering practicality through three integrated stages. 
    \textbf{Stage 1 (Skill Injection):} A Foundation LLM acts as a domain agent, translating natural language problem descriptions into a structured constraint set $\mathcal{S}=\{\mathcal{T}, \mathcal{O}, \mathcal{C}\}$, defining coordinate transformations, operator whitelists, and physical inequalities (e.g., Drucker stability). 
    \textbf{Stage 2 (Constrained Discovery):} The symbolic regression engine performs a genetic search governed by a physics-informed fitness function, balancing data fidelity with the penalties derived from $\mathcal{S}$. 
    \textbf{Stage 3 (FEM Verification):} Discovered models undergo rigorous finite element validation. Unlike classical empirical models which risk numerical divergence (Path A) or idealized models that lack flexibility (Path B), the EO-SR approach ensures the identification of constitutive laws that are both empirically accurate and numerically robust (Optimal Path).}
    \label{fig1}
\end{figure*}

The proposed Engineering-Oriented Symbolic Regression (EO-SR) framework addresses the fundamental disconnect between purely data-driven machine learning and the rigorous safety requirements of computational mechanics. Unlike traditional symbolic regression, which often functions as an unconstrained mathematical search engine prone to generating physically invalid expressions, EO-SR is architected as a modular, physics-governed discovery system. The general architecture of the proposed framework is illustrated in Figure \ref{fig1}. The framework is composed of three coupled modules: a Data Parser for standardization, a Skill Injector for physical knowledge integration, and a Symbolic Evolution Engine for constrained optimization.

The pipeline initiates with the \textbf{Data Parser}, which abstracts raw experimental inputs—potentially heterogeneous in unit or format—into a standardized tensor representation $\mathcal{D} = \{(\mathbf{x}_i, \mathbf{y}_i)\}_{i=1}^N$. This module ensures that subsequent discovery processes operate on a consistent numerical manifold, regardless of the underlying material domain.

Central to the architecture is the \textbf{Skill} Injector, an interface designed to bridge the gap between abstract physical laws and numerical regression. We formalize domain knowledge as a "Skill" tuple $\mathcal{S} = (\mathcal{T}, \mathcal{O}, \mathcal{C})$, which acts as a boundary condition for the search space. Here, $\mathcal{T}: \mathbb{R}^n \to \mathbb{R}^k$ represents a coordinate transformation function that maps raw inputs to a physically objective feature space (e.g., invariants or dimensionless numbers); $\mathcal{O}$ denotes the admissible operator set tailored to the domain's mathematical properties; and $\mathcal{C}$ represents a set of inequality constraints derived from conservation laws or thermodynamic principles.

The \textbf{Symbolic Evolution Engine} serves as the solver, executing a genetic search algorithm to identify functional forms $f: \mathbb{R}^k \to \mathbb{R}$ that map the transformed features to the target variable. Crucially, the search is not governed solely by goodness-of-fit but by a dynamic loss function $\mathcal{L}$ that enforces the injected skill. The global optimization problem is formulated as:

\begin{equation}
    \min_{f \in \mathcal{F}_{\mathcal{O}}} \left( \mathcal{L}_{fit}(f, \mathcal{D}) + \lambda_{reg} \Omega(f) + \sum_{c \in \mathcal{C}} \lambda_c \mathcal{P}(c, f) \right)
\end{equation}

where $\mathcal{F}_{\mathcal{O}}$ is the function space defined by the operator set $\mathcal{O}$, $\Omega(f)$ quantifies the model complexity (e.g., tree depth), and $\mathcal{P}(c, f)$ is a penalty functional that quantifies the violation of physical constraint $c$. By minimizing this composite objective, the framework ensures that the discovered models reside on the Pareto frontier of accuracy, simplicity, and physical consistency.

\subsection{Skill-Based Constraint Mechanism via Knowledge Agents}
A distinguishing feature of the EO-SR framework is its ability to automate the translation of high-level physical concepts into executable mathematical constraints, a process facilitated by a Large Language Model (LLM)—\textbf{specifically Gemini3-pro\cite{comanici2025gemini}}—acting as a Physics-Informed Agent. Rather than relying on rigid, hard-coded rules, the framework employs the Agent to synthesize the Skill tuple $\mathcal{S}$ through direct chain-of-thought inference based on its internal parametric knowledge.

Upon receiving a natural language domain descriptor (e.g., "Isotropic Hyperelasticity" or "Incompressible Fluid Dynamics"), the Agent deduces the necessary mathematical properties and generates the corresponding Python logic for the transformation $\mathcal{T}$, operators $\mathcal{O}$, and constraints $\mathcal{C}$. This eliminates the need for manual derivation or external retrieval databases, realizing a zero-shot transition from abstract concept to computational constraint.

For the constraint set $\mathcal{C}$, the Agent typically generates inequality conditions in the form $g(\mathbf{z}) \geq 0$, where $\mathbf{z}$ represents state variables or their derivatives. To integrate these inequalities into the gradient-based optimization of the SR engine, we formulate the penalty functional $\mathcal{P}$ using a rectified linear unit (ReLU) approach. For a generic constraint demanding positivity of a physical quantity $Q(f)$, the penalty is defined as:
\begin{equation}
    \mathcal{P}(Q, f) = \frac{1}{M} \sum_{j=1}^M \text{ReLU}\left( -Q(f(\mathbf{z}_j)) \right)^2
\end{equation}

where $M$ denotes the number of sampling points in the state space. This formulation ensures that physically valid regions contribute zero penalty, while violations induce a quadratic cost, guiding the evolutionary search back towards the physically feasible manifold. This mechanism allows the framework to enforce complex, derivative-dependent priors—such as thermodynamic stability or energy conservation—without requiring specific modifications to the underlying search algorithm.

\subsection{Instantiation: The Hyperelasticity Skill}
To validate the efficacy of the proposed framework, we tasked the Physics-Informed Agent with constructing a constitutive modeling workflow for rubber-like materials. Upon receiving the domain descriptor "Isotropic Hyperelasticity," the Agent instantiated a specific skill tuple, denoted as $\mathcal{S}_{hyper} = (\mathcal{T}_{iso}, \mathcal{O}_{mono}, \mathcal{C}_{stab})$, by retrieving and synthesizing fundamental principles from continuum mechanics.

The first component, the transformation logic $\mathcal{T}_{iso}$, was derived from the condition of material incompressibility ($\det \mathbf{F} = 1$) and the requirement of frame indifference. The Agent identified that the strain energy density $W$ for an isotropic hyperelastic solid must depend solely on the principal invariants of the right Cauchy-Green deformation tensor. To enforce the normalization condition that the strain energy vanishes in the undeformed state ($W(\mathbf{I}) = 0$), the Agent formulated the transformation logic $\mathcal{T}_{iso}$ to map experimental stretch ratios $\lambda$ into the shifted invariant space $\tilde{I}_1, \tilde{I}_2$ (where $\tilde{I}_k = I_k - 3$), strictly adhering to the incompressibility constraint ($J=1$). The specific mappings for Uniaxial Tension (UT), Equibiaxial Tension (ET), and Pure Shear (PS) are derived as:

\begin{equation}
\mathcal{T}_{iso}: \lambda \mapsto
\begin{cases}
\text{UT:} & \tilde{I}_1 = \lambda^2 + 2\lambda^{-1} - 3, \quad \tilde{I}_2 = 2\lambda + \lambda^{-2} - 3 \\
\text{ET:} & \tilde{I}_1 = 2\lambda^2 + \lambda^{-4} - 3, \quad \tilde{I}_2 = \lambda^4 + 2\lambda^{-2} - 3 \\
\text{PS:} & \tilde{I}_1 = \tilde{I}_2 = \lambda^2 + \lambda^{-2} - 2
\end{cases}
\end{equation}

This transformation ensures that the inputs to the symbolic regression engine are zero when the stretch ratio $\lambda=1$, thereby implicitly satisfying the zero-energy constraint without requiring additional penalty terms in the loss function.

For the operator set $\mathcal{O}_{mono}$, the Agent performed a context-aware pruning of the standard mathematical library. Recognizing that hyperelastic materials typically exhibit monotonic strain-hardening behavior, the Agent strictly excluded periodic functions (e.g., $\sin$, $\cos$) from the search space, as their oscillating nature contradicts the thermodynamic requirement of non-negative energy dissipation and can lead to non-physical softening regimes. Instead, the allowed operator set was optimized to include exponential (exp) and power (pow) functions, which are essential for capturing the severe stiffening associated with polymer chain locking at large deformations.

Finally, to guarantee thermodynamic consistency, the Agent formulated the constraint set $\mathcal{C}_{stab}$ based on the Drucker stability criterion. This criterion requires that the work done by external traction on any displacement increment must be non-negative. In the context of the strain energy potential, this implies global convexity with respect to the deformation invariants. The Agent implemented this as a hard constraint by computing the Hessian matrix of the discovered function $W(\tilde{I}_1, \tilde{I}_2)$ via automatic differentiation:

\begin{equation}
    \mathbf{H}_W = \begin{bmatrix} \frac{\partial^2 W}{\partial \tilde{I}_1^2} & \frac{\partial^2 W}{\partial \tilde{I}_1 \partial \tilde{I}_2} \\ \frac{\partial^2 W}{\partial \tilde{I}_2 \partial \tilde{I}_1} & \frac{\partial^2 W}{\partial \tilde{I}_2^2} \end{bmatrix} \succeq 0
\end{equation}

During the evolutionary search, any candidate expression that violates the positive semi-definiteness of $\mathbf{H}_W$ (i.e., $\det(\mathbf{H}_W) < 0$ or $\text{Tr}(\mathbf{H}_W) < 0$) within the domain of interest is penalized heavily. This mechanism acts as a "physical filter," ensuring that the EO-SR framework produces constitutive laws that are not only accurate in fitting experimental curves but also mathematically robust for downstream finite element analysis.

\subsection{Optimization and Discovery Process}
With the Hyperelasticity Skill ($\mathcal{S}_{hyper}$) fully instantiated, the Symbolic Evolution Engine is deployed to navigate the search space. The optimization process is configured as a high-throughput evolutionary strategy, utilizing a distributed island model to prevent premature convergence. For the specific showcase of rubber-like materials, the computational budget is set to 100 iterations across 30 parallel populations, driven by 4 concurrent processes. To maintain mathematical parsimony and prevent the generation of "bloated" expressions that are computationally expensive to evaluate, a hard constraint on expression complexity (maxsize) is set to 18 nodes. This configuration balances the breadth of exploration with the interpretability of the resulting expressions.



Upon completion of the evolutionary search, the engine outputs a Pareto frontier—a set of non-dominated solutions representing optimal trade-offs between fitting accuracy (MSE) and model complexity. While traditional approaches often rely on purely geometric heuristics (e.g., the "elbow method") to select a single "best" model, such metrics fail to capture the subtle mathematical properties critical for simulation readiness.

A fundamental challenge in symbolic regression is that penalty-based constraints (e.g., $\det \mathbf{H}_W \ge 0$) are only enforced at discrete sampling points. This leaves the discovered models vulnerable to interstitial instability—where small regions of negative curvature (non-convexity) may persist between sampled data points. In finite element analysis, even a localized violation of the Drucker stability criterion can lead to the loss of ellipticity and immediate solver divergence.

To bridge this gap, we introduce an Agent-Assisted Selection phase. The top-$k$ candidates are fed back to a Physics-Informed Agent equipped with a specialized Constitutive Modeling Skill. Unlike the SR engine, which treats formulas as computational graphs, the Agent leverages the domain expertise encoded in the Skill to audit the analytical form of the expressions. Specifically, it prioritizes candidates composed of inherently convex building blocks over high-order polynomials that may oscillate. For instance, guided by the Skill's instructions, the Agent identifies that rational terms of the form $\frac{A}{B-I_1}$ (where $A, B > 0$) guarantee strict convexity in the physical domain ($I_1 < B$), offering a superior stability guarantee compared to models that merely satisfy numerical checks on a grid.

Furthermore, the Agent favors expressions with clear physical interpretations—such as finite extensibility limits (singularities) or neo-Hookean linearity—over mathematically opaque constructs (e.g., nested roots) that may yield similar MSE values but lack theoretical justification. By integrating this semantic review, the framework ensures that the selected model is not merely a high-accuracy regression curve, but a structurally robust constitutive closure suitable for complex boundary value problems.

\section{Results}\label{sec3}

To demonstrate the efficacy and robustness of the proposed Engineering-Oriented Symbolic Regression (EO-SR) framework, we select the hyperelastic modeling of rubber-like materials as a representative showcase. This problem is chosen as a benchmark due to its inherent challenges: high non-linearity, incompressibility constraints, and the notorious difficulty in balancing fitting accuracy with numerical stability (Drucker stability).

Unlike recent data-driven approaches that rely on data-heavy full-field measurements (e.g., DIC-based VFM), our objective is to validate that the proposed framework can extract high-performance constitutive laws solely from standard industrial testing data (uniaxial and biaxial tension). The Treloar dataset, widely regarded as the "gold standard" in the field, serves as the ground truth for this validation.

In this section, we systematically evaluate the discovered models from four perspectives, following the workflow of our methodology:
\begin{itemize}
    \item \textbf{Model Discovery:} Analyzing the trade-off between complexity and accuracy via the Pareto frontier.
    \item \textbf{Fitting \& Generalization:} Verifying predictive capability on unseen deformation modes (e.g., pure shear).
    \item \textbf{Physical Consistency:} Rigorous checking of thermodynamic stability and physical limits.
    \item \textbf{Engineering Implementation:} Assessing numerical robustness in finite element simulations (FEM) under complex boundary conditions.
\end{itemize}

\subsection{Pareto Frontier and Model Discovery}
\begin{figure*}[htbp]
    \centering
    \includegraphics[width=0.95\linewidth]{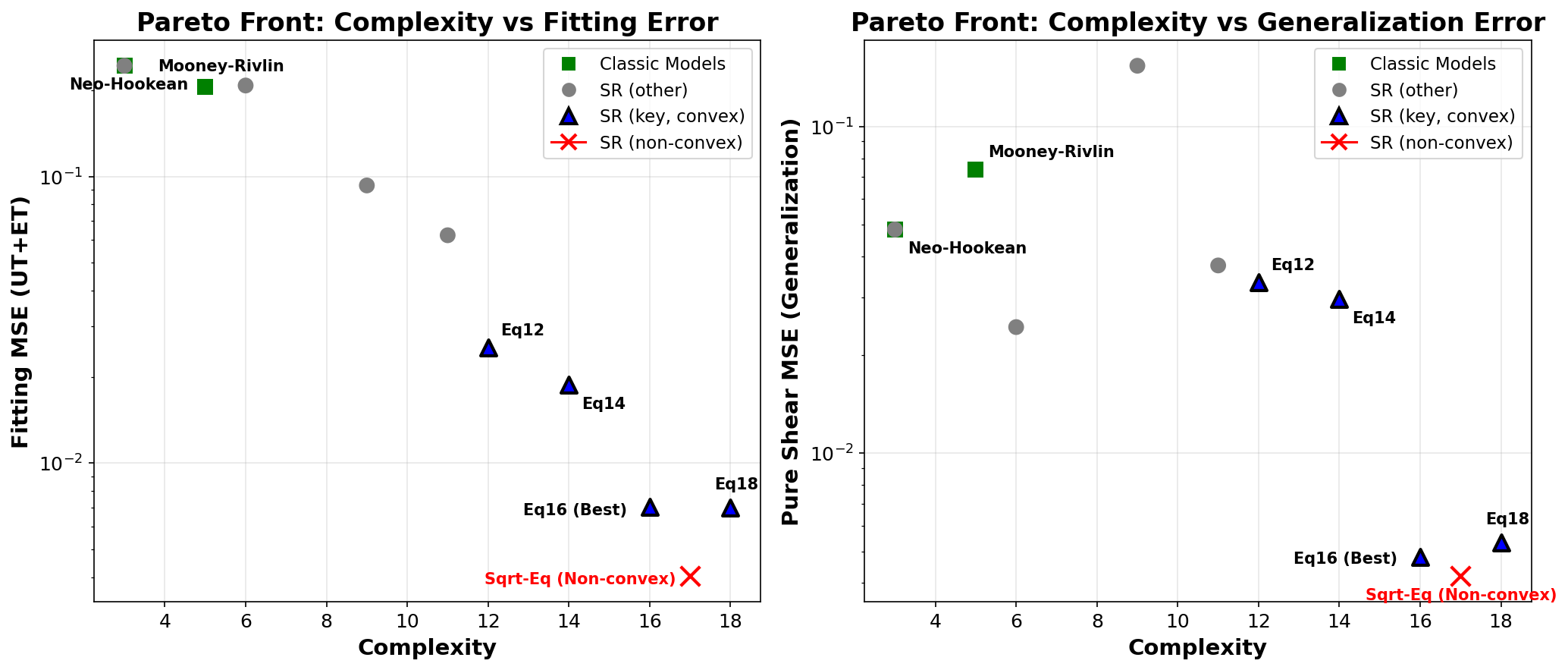}
    \caption{\textbf{Pareto frontier analysis of discovered constitutive models.} 
    The plots illustrate the trade-off between mathematical complexity (number of nodes) and accuracy. 
    \textbf{(Left)} Fitting MSE on training data (Uniaxial + Equibiaxial). 
    \textbf{(Right)} Generalization MSE on unseen Pure Shear data. 
    The proposed model (\textbf{Eq16}, blue triangle) identifies an optimal "elbow" point, achieving a balance between sparsity and precision. 
    Note that the unconstrained search found a lower-error solution (\textbf{Sqrt-Eq}, red cross), but it was rejected due to violation of the convexity (Drucker stability) constraint.}
    \label{fig2}
\end{figure*}

The trade-off between model complexity and predictive accuracy is visualized in the Pareto frontier (Figure \ref{fig2}). The symbolic regression (SR) algorithm, driven by the convexity-informed loss function, explored thousands of candidate expressions to identify the optimal constitutive models.

Model Selection Strategy The Pareto front reveals a distinct "elbow" pattern.
\begin{itemize}
    \item \textbf{Low Complexity Region (Complexity < 10):} Classic phenomenological models like Neo-Hookean (Complexity=3) and Mooney-Rivlin (Complexity=5) appear at the far left. While mathematically concise, they exhibit high fitting errors (MSE > 0.3), indicating underfitting for the highly non-linear Treloar dataset.
    \item \textbf{High Complexity Region (Complexity > 17):} As complexity increases, the fitting error drops significantly. However, unconstrained searches often lead to "bloated" mathematical structures. For instance, the non-convex solution (Sqrt-Eq) found at Complexity 17 achieves a very low MSE (~0.006) but utilizes a nested, non-physical structure:
    \begin{equation}
        W_{non-convex} = \sqrt{1.43 \left[ \exp\left(\sqrt{\sqrt{\exp(0.067 I_1)}} + 2.22\right) + \frac{I_1 + I_2 - 1.01}{1.87} \right]}
    \end{equation}
    This equation, despite its accuracy, violates the Drucker stability criterion (which will be discussed in subsequent chapters) and represents a typical case of overfitting to noise.
    \item \textbf{The Optimal Trade-off (The Proposed Model):} The most balanced solution is identified at Complexity 16 (Eq16). This model achieves a training MSE of ~0.008, comparable to high-order series expansions, but with a significantly sparser structure. It lies on the "sweet spot" of the Pareto curve, minimizing error without unnecessary mathematical inflation.
\end{itemize}

Structure of the Discovered Model The best-performing model discovered by our framework (Eq16) is given by:
\begin{equation}
    W(\tilde{I}_1, \tilde{I}_2) = 0.031(3.75 \tilde{I}_1 + \tilde{I}_2) + \frac{\tilde{I}_1}{77.9 - 1.05 \tilde{I}_1}
\end{equation}
where $\tilde{I}_1 = I_1 - 3$ and $\tilde{I}_2 = I_2 - 3$.

This formula is not merely a statistical fit but a hybrid physical model automatically synthesized by the algorithm: 
\begin{itemize}
    \item Linear Term ($0.116 \tilde{I}_1 + 0.031 \tilde{I}_2$): The first part corresponds exactly to the classical Mooney-Rivlin model, capturing the linear elasticity at small to moderate deformations. The discovered parameter ratio $C_{10}/C_{01} \approx 3.75$ aligns well with typical rubber-like materials.
    \item Rational Term ($\frac{\tilde{I}_1}{77.9 - 1.05 \tilde{I}_1}$): The second part introduces a rational function with a singularity. As $\tilde{I}_1$ approaches $\approx 74.2$ (corresponding to a stretch $\lambda \approx 8.8$), the strain energy tends to infinity. This term acts as a Gent-like locking function, physically representing the finite extensibility of polymer chains.
\end{itemize}

Table\ref{tab1} summarizes the top candidate formulas on the Pareto frontier. It is worth noting that while Sqrt-Eq offers a marginally lower error, it introduces additional terms without a clear physical justification. Therefore, Eq16 is selected as the optimal constitutive law for further validation

\begin{table}[htbp]
    \centering
    \caption{Summary of top candidate constitutive models identified on the Pareto frontier. The proposed model (Eq16) achieves the best trade-off between complexity and physical capture.}
    \label{tab1}
    \renewcommand{\arraystretch}{1.2} 
    \begin{tabular}{l c c l l}
        \toprule
        \textbf{Model ID} & \textbf{Comp.} & \textbf{MSE (Fitting)} & \textbf{Mathematical Structure} & \textbf{Physical Remark} \\
        \midrule
        Mooney-Rivlin & 5 & $\sim 0.35$ & $C_{10}\tilde{I}_1 + C_{01}\tilde{I}_2$ & Underfits large deformations \\
        
        Eq12 & 12 & $\sim 0.015$ & $\frac{\tilde{I}_1}{c_1 \tilde{I}_1 + c_2}$ & Misses shear components \\
        
        \textbf{Eq16 (Best)} & \textbf{16} & \textbf{$\sim 0.008$} & \textbf{$W_{MR} + \frac{\tilde{I}_1}{\tilde{I}_{lim} - \tilde{I}_1}$} & \textbf{Captures locking \& shear} \\
        
        Sqrt-Eq & 17 & $\sim 0.006$ & $\sqrt{\exp(\dots) + \text{Poly}}$ & \textbf{Non-convex} (Unstable) \\
        \bottomrule
    \end{tabular}
    \vspace{0.2cm} 
    
    \footnotesize
    \raggedright
    \textit{Note:} $Comp.$ denotes the complexity score (number of nodes in the expression tree). $W_{MR}$ represents the linear Mooney-Rivlin terms. $\tilde{I}_1 = I_1 - 3$ and $\tilde{I}_2 = I_2 - 3$ are the shifted invariants.
\end{table}

\subsection{Fitting Accuracy and Generalization Capability}
\begin{figure*}[htbp]
    \centering
    \includegraphics[width=\textwidth]{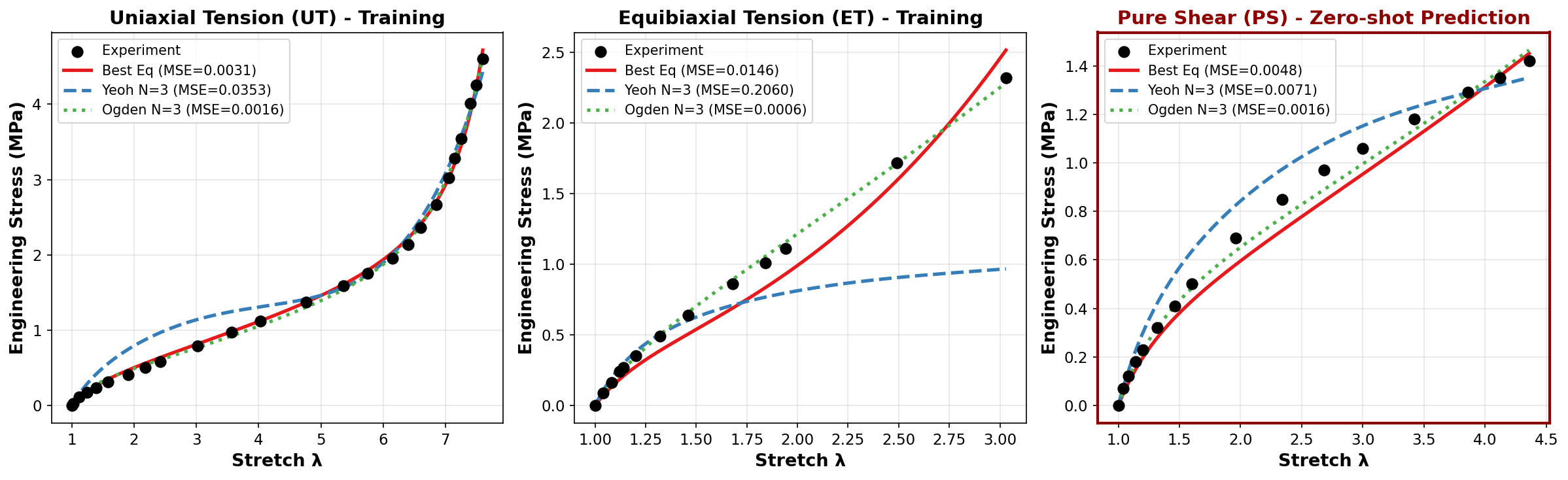}
    \caption{\textbf{Performance comparison of the discovered model against industry benchmarks.} 
    The stress-stretch responses are evaluated under: 
    \textbf{(Left)} Uniaxial Tension (Training), 
    \textbf{(Center)} Equibiaxial Tension (Training), and 
    \textbf{(Right)} Pure Shear (Zero-shot Prediction). 
    The proposed SR model (\textbf{Eq16}, solid red line) demonstrates superior multi-axial capability compared to the Yeoh model (N=3, dashed blue), which fails to capture the equibiaxial response. 
    Crucially, in the pure shear regime (not used for calibration), the SR model accurately predicts the experimental data ($MSE \approx 0.0048$), verifying its physical generalization capability compared to polynomial overfitting.}
    \label{fig3}
\end{figure*}

To evaluate the predictive fidelity of the discovered constitutive law (Eq16), we compared its performance against the experimental data and two industry-standard benchmarks: the Yeoh model (N=3) (representing polynomial expansions) and the Ogden model (N=3) (representing high-order series expansions). Figure \ref{fig3} presents the stress-stretch response across three deformation modes: Uniaxial Tension (UT), Equibiaxial Tension (ET), and Pure Shear (PS).

Calibration Performance (UT and ET) The model was calibrated solely using UT and ET datasets. As shown in the left and center panels of Figure \ref{fig3}, the proposed SR model (solid red line) demonstrates excellent agreement with the experimental data.
\begin{itemize}
    \item \textbf{Uniaxial Tension:} The model captures the entire "S-shaped" curve, including the initial softening and the stiffening at large strains ($\lambda > 6$), with an MSE of $0.0031$.
    \item \textbf{Equibiaxial Tension:} A critical advantage of the discovered model is observed here. While the Yeoh model (blue dashed line) significantly underestimates the stress at larger strains ($MSE \approx 0.2060$) due to its reliance solely on the first invariant $I_1$, the SR model accurately tracks the biaxial hardening behavior ($MSE \approx 0.0146$). This improvement is attributed to the inclusion of the linear $\tilde{I}_2$ term in Eq16, which effectively captures the deformation-mode dependence that $I_1$-based models inherently miss.
    \item \textbf{Benchmark Comparison:} It is noted that the Ogden model (green dotted line) achieves the lowest training error. This is expected, as the Ogden model utilizes 6 independent parameters (compared to 4 in Eq16) and is mathematically a series expansion capable of adhering closely to data points. However, the SR model achieves comparable accuracy with a significantly sparser and more interpretable structure.
\end{itemize}

\textbf{Zero-Shot Generalization (Pure Shear)} The robustness of a constitutive model is best tested against deformation modes not used in calibration. The right panel of Figure \ref{fig3} illustrates the "zero-shot" prediction for Pure Shear (PS).

The robustness of the discovered constitutive law is most rigorously tested in the pure shear regime, a deformation mode completely absent from the symbolic search process. Despite this "zero-shot" condition, the SR model yields a remarkably accurate prediction ($MSE \approx 0.0048$), with the generated curve passing almost directly through the experimental points. This performance highlights a critical divergence in modeling approaches: while the Yeoh model exhibits an artificial stiffening response that overestimates stress—symptomatic of polynomial overfitting—the SR model correctly interpolates the material response between the uniaxial and biaxial limits. This capability is strictly attributed to its hybrid structure, which balances Mooney-Rivlin linear terms with the rational locking term, confirming that the framework has identified a constitutive law with genuine physical validity rather than merely overfitting specific loading paths.

\subsection{Thermodynamic Consistency and Stability Analysis}
While fitting accuracy describes how well a model replicates past observations, thermodynamic consistency determines whether the model is physically permissible and numerically safe for future predictions. Therefore, we rigorously evaluate the discovered constitutive law (Eq16) against the Drucker stability criterion, ensuring it meets the stringent requirements for finite element implementation.

\begin{figure}[htbp]
    \centering
    \includegraphics[width=\linewidth]{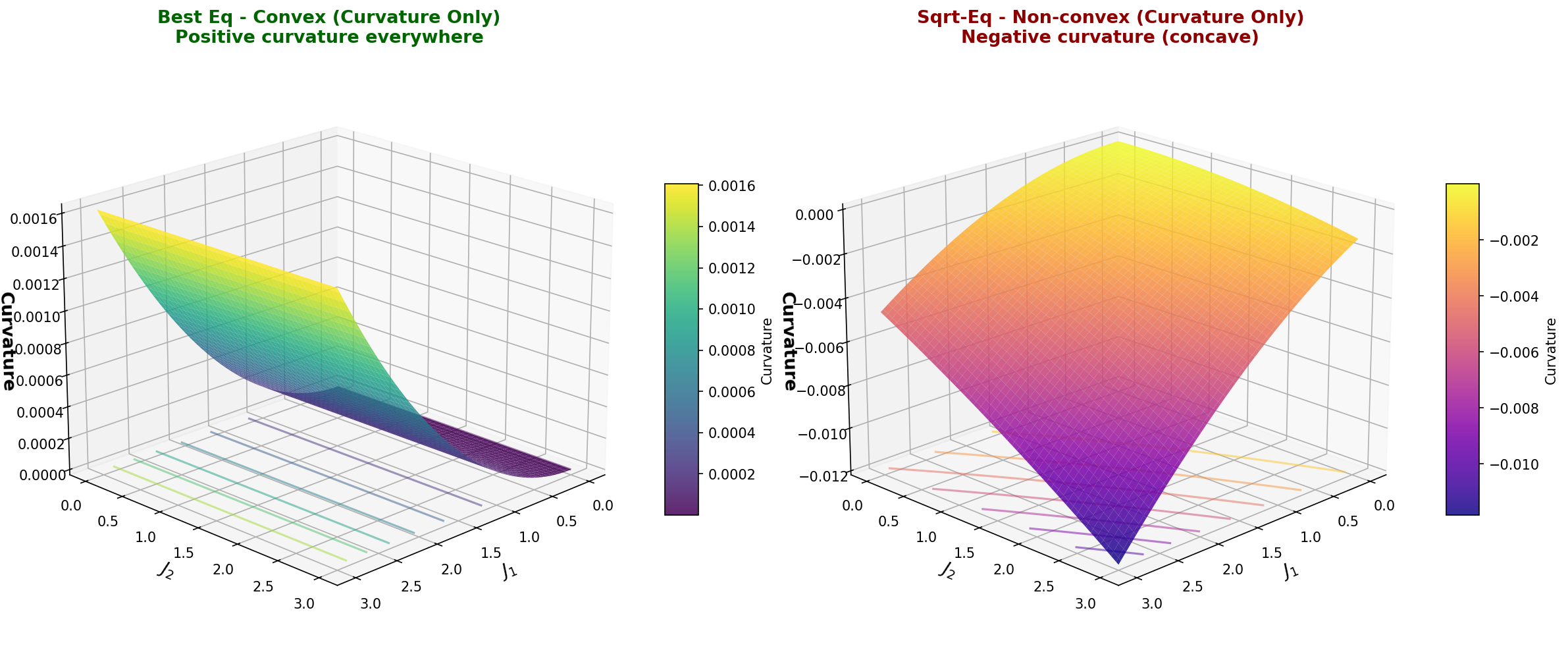}
    \caption{\textbf{Visualization of thermodynamic consistency via energy landscape curvature.} 
    \textbf{(Left)} The proposed convex model (\textbf{Eq16}) exhibits a positive-definite Hessian (smooth bowl shape) across the entire invariant space, guaranteeing unconditional stability. 
    \textbf{(Right)} The unconstrained SR solution (\textbf{Sqrt-Eq}) reveals regions of negative curvature (concavity), indicating physical instability despite high fitting accuracy. 
    This comparison highlights the critical role of physics-informed constraints in filtering out mathematical artifacts.}
    \label{fig4}
\end{figure}

\textbf{Global Convexity and Energy Landscape}
The primary condition for hyperelastic stability is that the strain energy density function $W(\tilde{I}_1, \tilde{I}_2)$ must be strictly convex. Figure \ref{fig4} visualizes the curvature of the energy surface in the invariant phase space to assess this property. The proposed model (Eq16) exhibits a smooth, "convex bowl" shape with positive curvature across the entire domain, geometrically guaranteeing that the Hessian matrix remains positive semi-definite. This property ensures that for any deformation path, the material response remains unique and stable ($\det(\mathbf{H}) \geq 0$). In stark contrast, the unconstrained solution (Sqrt-Eq), despite its lower fitting error, reveals distinct regions of negative curvature (saddle points). Such non-convexity poses a severe risk in computational mechanics, as it implies that under specific complex loading conditions, the global stiffness matrix could become singular, leading to immediate non-convergence in solvers. This comparison validates the necessity of the "convexity-informed" loss function embedded in our framework for filtering out mathematical artifacts.

\begin{figure}[htbp]
    \centering
    \includegraphics[width=\linewidth]{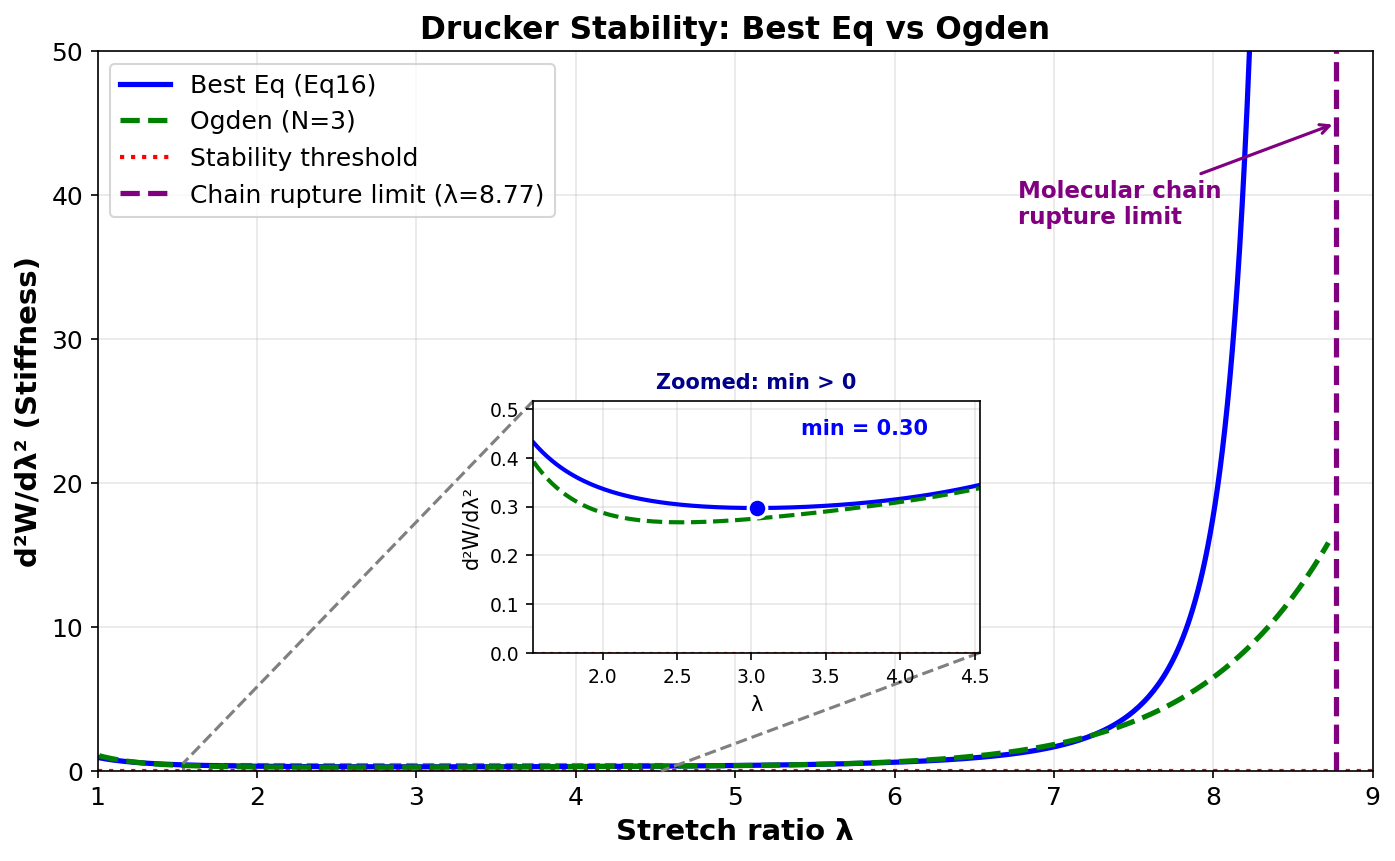}
    \caption{\textbf{Analysis of Drucker stability and finite extensibility limit.} 
    The plot shows the evolution of tangent stiffness ($d^2W/d\lambda^2$) under uniaxial tension. 
    \textbf{Inset:} The proposed model maintains strict positivity (min stiffness $\approx 0.30$ MPa), satisfying the stability prerequisite. 
    \textbf{Main Plot:} At large deformations, the Ogden model (dashed green) predicts unbounded polynomial growth. In contrast, the proposed SR model (solid blue) successfully identifies the molecular chain rupture limit, exhibiting an asymptotic stiffness surge at $\lambda \approx 8.77$. This physical "locking" behavior prevents non-physical infinite stretching in simulation.}
    \label{fig5}
\end{figure}

\textbf{Tangent Stiffness and Finite Extensibility}
To further assess physical validity at extreme deformations, we analyzed the uniaxial tangent stiffness modulus ($d^2W/d\lambda^2$), which serves as a one-dimensional indicator of Drucker stability. As illustrated in Figure \ref{fig5}, the stiffness of the proposed model maintains strict positivity (min $\approx 0.30$ MPa) throughout the deformation range, effectively preventing non-physical softening or numerical oscillations. More importantly, a significant divergence in physical fidelity is observed in the large deformation regime ($\lambda > 7$). The Ogden model follows a polynomial growth pattern ($\sim \lambda^\alpha$), suggesting that the material can stretch indefinitely without limit. Conversely, the SR model correctly predicts a vertical asymptote at $\lambda \approx 8.77$. This behavior, intrinsic to the discovered rational term ($\frac{\tilde{I}_1}{77.9 - 1.05 \tilde{I}_1}$), phenomenologically reproduces the finite extensibility limit (molecular chain rupture). This feature acts as a physical "safety barrier," preventing elements from inverting under extreme loads where polynomial models would fail to provide sufficient resistance.

In summary, the SR-discovered model achieves a rare balance in constitutive modeling: it possesses the unconditional mathematical stability required for numerical convergence—similar to the Neo-Hookean model—while simultaneously capturing the complex stiffening physics of real polymers, a capability often lacking in standard polynomial expansions.

\subsection{Finite Element Implementation and Engineering Robustness}

The ultimate validation of a constitutive model lies in its numerical robustness under complex, inhomogeneous boundary conditions. We implemented the discovered constitutive law (Eq16) as a UMAT in Abaqus/Standard and subjected it to the large deformation of a double-edge notched tensile specimen. This setup creates a rigorous test environment where tension, shear, and transverse compression co-exist.

\begin{figure*}[htbp]
    \centering
    \includegraphics[width=\textwidth]{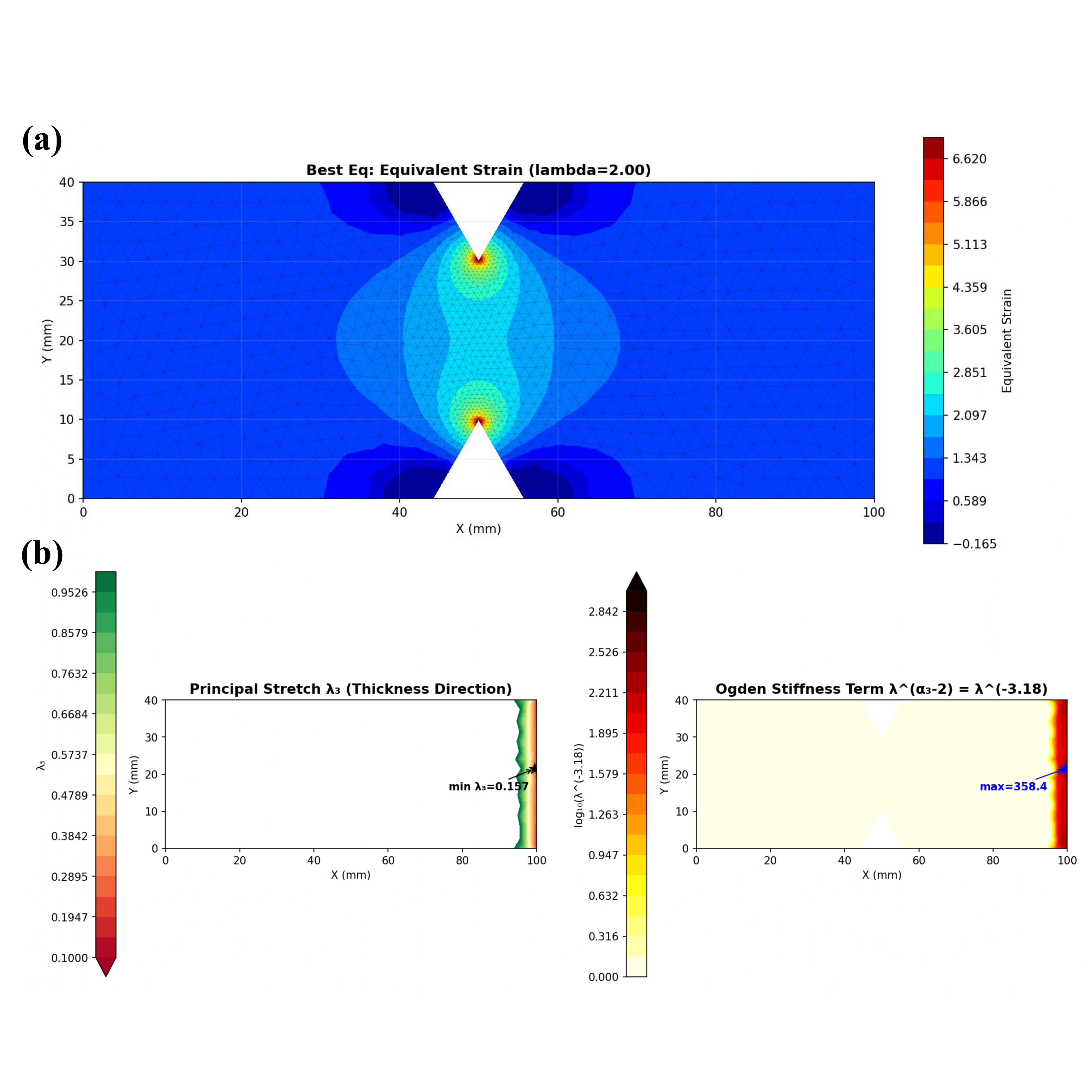}
    \caption{\textbf{Finite element validation and failure mechanism analysis on a double-edge notched specimen.} 
    \textbf{(Top) Robustness of the proposed SR model:} The simulation successfully reaches the target large deformation ($\lambda_{global} = 3.0$). The equivalent strain contour remains smooth and continuous, demonstrating the numerical stability provided by the convexity-informed rational formulation which inherently prevents stiffness singularities.
    \textbf{(Bottom) Mechanism of Ogden model divergence:} Forensic analysis of the boundary elements where the benchmark simulation (Ogden N=3) failed. Due to material incompressibility, the clamped boundary induces severe transverse compression ($\lambda_3 \approx 0.157$). The Ogden model, fitted with a negative exponent ($\alpha_3 \approx -3.18$) to minimize tensile error, exhibits a numerical singularity in this compressive regime. The stiffness term $\lambda^{\alpha_3-2}$ explodes to $\approx 358$, causing the tangent stiffness matrix to become ill-conditioned. This confirms that data-driven models without global convexity constraints are prone to failure in multi-axial deformation modes.}
    \label{fig6}
\end{figure*}

\textbf{Simulation Success of the Proposed Model} Figure \ref{fig6}(a) presents the deformation state of the SR model at maximum displacement. The model successfully handles the large stress concentrations at the notch tips without convergence difficulties. The equivalent strain field is smooth, and the mesh maintains good quality even under severe distortion. This success is attributed to the rational locking term in Eq16, which provides a physical "hardening wall," ensuring a positive-definite stiffness matrix across all deformation regimes.

\textbf{Forensic Analysis of Ogden Failure} In contrast, the simulation using the Ogden model (N=3) failed to converge. Rather than simply dismissing this as a "numerical error," we analyzed the mathematical root cause in Figure \ref{fig6}(b).

\begin{itemize}
    \item \textbf{The Hidden Trap:} While the specimen is pulled in tension ($X$-direction), the incompressibility constraint ($\det \mathbf{F} = 1$) forces the material to contract significantly in the thickness direction ($Z$-direction). At the clamped boundaries, the principal stretch drops to $\lambda_3 \approx 0.157$ (highly compressed).
    \item \textbf{Mathematical Singularity:} The Ogden parameters, optimized solely for fitting tensile data ($\lambda > 1$), included a large negative exponent ($\alpha_3 \approx -3.18$). As visualized in the heatmap of Figure \ref{fig6}(Bottom), this term causes the stiffness component proportional to $\lambda_3^{\alpha_3-2}$ to explode exponentially (reaching values > 350) as $\lambda_3 \to 0$.
    \item \textbf{Consequence:} This localized stiffness spike creates an ill-conditioned Jacobian matrix that the Newton-Raphson solver cannot resolve. This finding highlights a critical insight: models discovered via unconstrained fitting are liable to contain "blind spots" (e.g., compression instability) that are only triggered in multi-axial boundary value problems.
\end{itemize}

\section{Discussion}
\subsection{Beyond Isotropy: Generalizing Logical Constraints}\label{sec4.1}

To underscore the methodological universality of the Agent-Skill framework beyond the initial demonstration on isotropic hyperelasticity, we present a conceptual case study on fiber-reinforced soft tissues (e.g., arterial walls). This scenario is particularly illustrative because it exposes a fundamental limitation of traditional symbolic regression: the inability to spontaneously discover logical branching constraints without explicit human intervention.

\textbf{The Challenge: Logical Discontinuities}
Traditional SR engines, driven purely by MSE, treat constitutive laws as smooth, continuous functions. However, fiber-reinforced tissues exhibit a "switch-like" behavior known as Tension-Compression Asymmetry: fibers stiffen under tension but buckle (contribute zero stiffness) under compression. A standard SR search typically converges to smooth polynomials like $k(I_4-1)^2$, which effectively minimizes error in the tensile regime but physically implies that fibers resist compression—a catastrophic violation of the material physics.

\textbf{The Agent-Skill Solution}
To solve this, our framework leverages the constraint set $\mathcal{S}=\{\mathcal{T}, \mathcal{O}, \mathcal{C}\}$  to guide the discovery process. 
\begin{itemize}
    \item \textbf{$\mathcal{T}$: Additive Decomposition} $\mathcal{T}$ defines the mandatory functional architecture. The Agent enforces an additive split between the isotropic matrix and the anisotropic fibers:
\begin{equation}
W(I_1, I_4) = W_{iso}(I_1) + W_{f}(I_4)    
\end{equation}
Here, $W_{iso}(I_1)$ retains the isotropic formulations (e.g., Neo-Hookean, Yeoh) validated in Section \ref{sec3}, while $W_{f}(I_4)$ introduces the direction-dependent reinforcement, where $I_4 = \lambda_{fiber}^2$ represents the squared fiber stretch.
    \item \textbf{$\mathcal{O}$: Semantic Filtering} The operator library is effectively curated to satisfy $\mathcal{C}$. Periodic functions (e.g., $\sin, \cos$) are excluded due to the monotonic nature of tissue stiffening. Crucially, the library is augmented with conditional operators (e.g., Macaulay brackets $\langle x \rangle = \max(0, x)$) to structurally enable the "switching" behavior required by the buckling constraint.
    \item \textbf{$\mathcal{C}$: Asymmetry and Convexity} $\mathcal{C}$ imposes strict inequalities to ensure thermodynamic consistency. The fiber term must be active only in tension ($I_4 > 1$) and exhibit strict convexity to ensure stability:
\begin{equation}
    W_f = 0 \quad \forall I_4 \le 1; \qquad \frac{\partial^2 W_f}{\partial I_4^2} > 0 \quad \forall I_4 > 1
\end{equation}
\end{itemize}

Guided by this protocol, the Agent steers the symbolic search space toward structurally valid candidates. Ideally, this process isolates models akin to the Holzapfel-Gasser-Ogden (HGO) prototype:
\begin{equation}
W = C_{10}(I_1 - 3) + \frac{k_1}{2k_2} \left( \exp \left[ k_2 \langle I_4 - 1 \rangle^2 \right] - 1 \right)
\end{equation}
Here, the framework enables the Agent to identify that the exponential term satisfies the strict convexity required by $\mathcal{C}$, while the Macaulay brackets $\langle \cdot \rangle$ fulfill the logical switching mandated by $\mathcal{O}$. This mechanism not only recovers standard formulations but also allows for the discovery of novel, physically interpretable augmentations that strictly adhere to thermodynamic constraints.

\textbf{Experimental Scope}
It is worth noting that validating such anisotropic models requires planar biaxial testing to decouple matrix and fiber properties, an experimental complexity beyond the scope of this work. However, this conceptual demonstration highlights the framework's unique role as a semantic compiler, capable of translating high-level physical logic into structural constraints that purely numerical methods cannot grasp.

\subsection{Limitations and Future Directions}
While the EO-SR framework demonstrates a significant leap in automating the discovery of constitutive laws, several challenges remain to be addressed in future iterations.

\textbf{Experimental Scope and Data Heterogeneity}
Regarding the generalization to anisotropic tissues discussed in Section \ref{sec4.1}, a practical challenge lies in data richness rather than sensor resolution. Unlike high-barrier approaches that demand full-field measurements (e.g., DIC), our framework is designed to operate on standard, accessible testing data. However, for fiber-reinforced materials, standard uniaxial tests are insufficient to decouple the matrix ($W_{iso}$) from the fiber ($W_f$) contributions. Future work will not require upgrading to expensive optical setups, but rather utilizing Planar Biaxial Testers—common in characterizing soft tissues—to generate multi-ratio stress-stretch curves (e.g., $1:1, 1:0.5, 0.5:1$ loading protocols). This ensures that the Agent has sufficient phenomenological data to distinguish direction-dependent invariants without abandoning the "low-barrier" philosophy of the EO-SR framework.

\textbf{Computational Overhead and Agent Latency}
While the Symbolic Regression engine is highly efficient, the "Skill Injection" and "Agent-Assisted Selection" phases involve latency associated with Large Language Model inference. Although our architecture strategically places the Agent outside the evolutionary loop (acting as a pre-processor for constraints and a post-processor for selection) to minimize cost, the token consumption for processing complex constraint logic can be non-negligible. As Foundation Models evolve towards domain-distilled checkpoints, we anticipate deploying localized "Mechanics-Agents," significantly reducing latency and addressing privacy concerns regarding proprietary material data.

\section{Conclusion}
In this study, we introduced Engineering-Oriented Symbolic Regression (EO-SR), a generalized framework designed to resolve the epistemic uncertainty in computational physics by automating the discovery of physically valid closure relations. By integrating Large Language Models (LLMs) as "Physics-Informed Agents," we demonstrated that the abstract principles governing material behavior can be translated into executable constraints, effectively transforming symbolic regression from a mathematical curve-fitter into a physics-governed discovery engine.

The efficacy of this framework was demonstrated through the rigorous "torture test" of hyperelastic constitutive modeling, yielding three broader implications for the field:
\begin{enumerate}
    \item \textbf{Automated Alignment with Physical Laws:} The successful zero-shot generation of Drucker stability constraints confirms that LLMs can bridge the semantic gap between high-level physical theory and low-level optimization logic. This capability is domain-agnostic, paving the way for automated constraint injection in fields ranging from fluid mechanics (e.g., enforcing energy cascades) to thermodynamics (e.g., entropy production).
    \item \textbf{Discovery of "Simulation-Ready" Mathematics:} Our findings highlight that numerical robustness is not merely an implementation detail but a fundamental criterion for discovery. The framework’s rejection of "high-accuracy but non-convex" models (like the Sqrt-Eq) in favor of the unconditionally stable Rational-Eq (Eq16) underscores the necessity of embedding solver-relevant constraints directly into the discovery loop.
    \item \textbf{Generalizability of the "Skill" Architecture:} While this work focused on hyperelasticity, the modular design of the "Skill Injector" allows for seamless extension to other complex systems. Future work will explore the instantiation of "Plasticity Skills" (yield surfaces) and "Anisotropy Skills" (fiber reinforcement), further validating EO-SR as a universal tool for constitutive discovery.
\end{enumerate}

Ultimately, this work suggests a new paradigm for ``AI for Science,'' where foundation models serve not merely as coding assistants, but as the guardians of theoretical consistency. By enforcing fundamental axioms—such as frame indifference and thermodynamic stability—this framework elevates AI-discovered models from mere empirical regressions to mathematically rigorous constitutive closures, bridging the divide between data-driven description and predictive physical realism.

\bibliographystyle{unsrt}  


\bibliography{references}

\end{document}